\documentclass[pdflatex,sn-apa]{sn-jnl}

\usepackage{graphicx}%
\usepackage{multirow}%
\usepackage{amsmath,amssymb,amsfonts}%
\usepackage{amsthm}%
\usepackage{mathrsfs}%
\usepackage[title]{appendix}%
\usepackage{xcolor}%
\usepackage{textcomp}%
\usepackage{manyfoot}%
\usepackage{booktabs}%
\usepackage{algorithm}%
\usepackage{algorithmicx}%
\usepackage{algpseudocode}%
\usepackage{listings}
\usepackage{array}       

\begin{document}
\title[Patterns and Purposes]{Patterns and Purposes: A Cross-Journal Analysis of AI Tool Usage in Academic Writing}

\author*[1]{\fnm{Ziyang} \sur{Xu}}\email{ziyang.xu-1@ou.edu}

\affil*[1]{\orgdiv{School of Library and Information Studies}, 
\orgname{University of Oklahoma}, 
\orgaddress{\street{Bizzell Library, Room 120, 401 West Brooks}, 
\city{Norman}, 
\postcode{73019-6032}, 
\state{Oklahoma}, 
\country{United States}}}

\abstract{This study investigates the use of AI tools in academic writing through an analysis of AI usage declarations in  journals. Using a mixed-methods approach combining content analysis, statistical analysis, and text mining, this study analyzed 135 AI declarations from 8633 articles across 27 categories. Results show that ChatGPT dominates academic writing assistance (73.3\% usage). Crucially, the primary purposes of AI integration are heavily concentrated on lower-level cognitive tasks, specifically improving readability (57.8\%) and grammar checking (19.3\%). Furthermore, statistical analysis indicates a highly significant association between team composition and AI-use purposes ($p = 0.0008$), highlighting international teams' specific reliance on grammar assistance, while no significant association was found regarding authors' native-speaker status ($p = 0.2359$).  These findings provide insights for journal policy development and understanding the evolving role of AI in academic writing.
}




\maketitle

\section{Introduction}\label{sec1}

With the rise of generative AI, an increasing number of AI tools are being introduced into the academic field. Generative AI models, led by ChatGPT, assist in drafting manuscripts, thereby improving the writing efficiency of researchers and authors \citep{thomas2023impact}. These features can assist in various stages of the academic writing process, from idea generation, literature review writing, and data analysis,and even proposal creation \citep{abd2023artificial}. AI's ability to generate coherent and contextually relevant text has led to its adoption in academic settings, where it is used to streamline the writing process and support researchers in organizing their ideas \citep{hsu2023can}.

However, this has also sparked concerns about the potential for AI to undermine critical thinking and creativity. The key issues surrounding AI in academic publishing involve authorship and attribution of AI-generated content, data privacy concerns, and the appropriate scope for AI application \citep{mrabet2023chatgpt,storey2023ai}. In academic writing, \citet{dubose2023ai} highlight the need for responsible and careful use of AI tools. \citet{caprioglio2023fake} advocate for the creation of clear guidelines to ensure the accuracy and reliability of AI-generated content. 

In response to this challenge, academic publishers have revised their editorial policies and author guidelines to offer clearer instructions on disclosing the use of generative AI in academic research. The \textit{Science} series of journals has consistently required authors to sign a license certificate, ensuring that their published papers are original works \citep{thorp2023chatgpt}. According to \citet{ganjavi2024publishers}, 24\%\ of the top 100 largest publishers offer guidelines for the use of generative AI. The most notable among them is Elsevier, which offers a specific AI usage disclosure template and recommends incorporating it into a new, dedicated section of the manuscript. This template requires authors to disclose whether they used generative artificial intelligence during the writing process, specify the type of AI used, and explain its primary purpose.

The current discussion on AI in academic writing primarily centers on ethical controversies, potential impacts, and related issues. However, there is still limited research on the specific purposes behind authors' use of AI in academic writing, as well as how this relates to their backgrounds and the journals in which they publish. This study aims to collect and analyze the 'Declaration of Generative AI and AI-assisted technologies in the writing process' from selected Elsevier open-access journal sources . Through content analysis, association analysis, and text mining techniques, the research will explore the key components of AI usage declarations and investigate whether there are consistencies or variations in these declarations across different author teams and their backgrounds. The goal is to offer insights that can inform academic journal policies and broaden the discussion on the ethical and social implications of AI usage.

\section{Literature Review}

\subsection{Benefits of GAI for scholarly publishing}

Before investigating the role of generative artificial intelligence (GAI) in academic writing, it is essential to understand why researchers are drawn to using GAI in the first place. Most scholars hold a balanced perspective on the impact of GAI on academic writing. They believe that GAI can enhance writing efficiency, provide new ideas, correct grammatical structures and spelling errors, and offer personalized guidance \citep{lund2023chatting,kasneci2023chat,DWIVEDI2023102642}. Some studies explore the assistive value of ChatGPT in writing, demonstrating potential both for peer-reviewed literature reviews and for content analysis of published literature \citep{dergaa2023from,rahman2023chatgpt,AydinKaraarslan2023ChatGPT}. To validate ChatGPT's exceptional writing abilities, two scientists, with the assistance of ChatGPT, completed a research paper in under an hour. The study found that ChatGPT could not only handle data processing but also polish the initial draft \citep{conroy2023scientists}.

In empirical research, \citet{fyfe2023how} conducted a study involving 20 university students across different academic levels, investigating their experiences using GPT-2 to complete their final course papers. Eighty-seven percent of the students reported that collaborative writing with GPT was more manageable than completing the work independently, as it provided new arguments and ideas. Science published a controlled experimental study involving professionals from various fields, showing that ChatGPT is especially beneficial for individuals with weaker writing skills \citep{noy2023experimental}. The results indicated that ChatGPT can elevate their writing abilities to a level close to that of proficient writers.

\subsection{Ethical Concerns and Policies for AI in Academic Publishing}
As early as 2020, two initiatives for designing AI intervention reports were introduced in the medical field: CONSORT-AI (Consolidated Standards of Reporting Trials for Artificial Intelligence) and SPIRIT-AI \citep{Liu2020,CruzRivera2020}. In February 2023, the Committee on Publication Ethics (COPE) issued a position statement on the use of AI tools in research publications, emphasizing that AI tools cannot be listed as co-authors and providing guidelines on how to disclose the use of AI \citep{COPE2023}. This statement was quickly endorsed by journals and editorial associations such as the International Committee of Medical Journal Editors (ICMJE), the Journal of the American Medical Association (JAMA), and the World Association of Medical Editors (WAME) \citep{ICMJE2023,Flanagin2023,WAME2023}. The introduction of such policies is precisely due to the growing concerns among scholars about the ethical issues surrounding the use of generative AI in academic publishing \citep{Anderson2023}.  Large language models can produce a phenomenon known as "hallucination," where the generated text appears coherent and meaningful but is actually fabricated. In addition, since GPT-3 is trained on large web-based datasets, it inevitably inherits racial, gender, and ethnic biases during the training process \citep{Basta2019,Founta2018,Hutchinson2020}. A recent study investigating the academic writing capabilities of mainstream generative AI systems found that AI‑generated texts tend to exhibit poor readability and relatively high similarity rates \citep{Aydin2026GenerativeAIComparison}. These biases can be propagated through the publication of academic articles, further amplifying their impact within the scholarly community. 
\subsection{Impetus for the current study}
However, these concerns have largely remained in the realm of theoretical discussions. There is a lack of empirical investigation into the actual purposes behind researchers' use of generative AI in academic writing and the potential ethical and academic impacts it may cause. While much has been written about how AI is being integrated into academia, there is a gap in understanding the reasons for AI use in writing. To bridge this gap, it is crucial to move beyond theoretical debates and examine authors' self-disclosed practices. By systematically analyzing AI usage declarations in published manuscripts, this study aims to uncover the tangible patterns of human-AI collaboration in academic publishing.

\section{Research Questions}

Guided by the aforementioned need to understand practical AI adoption patterns, this paper aims to address the following research questions:\begin{enumerate}
\renewcommand{\labelenumi}{(\theenumi)}  
\item What types of AI tools are currently being used by researchers to assist in academic writing?
\item What specific tasks are AI tools primarily used to accomplish?
\item Are declared AI-use purposes associated with native-speaker status and international-team status?
\end{enumerate}

\section{Methodology}
\subsection{Journal selection and data acquisition}

The data collection scope of this study is as follows. Based on the Scopus database (\texttt{https://www.scopus.com/home.uri}), this study focuses on journals classified within the 27 major subject categories defined by Scopus. For each category, the highest-ranked journal was selected based on the CiteScore metric. To facilitate the analysis of AI usage declarations,the scope was further restricted to open-access journals published by Elsevier.  One reason for selecting Elsevier journals as the research context is that Elsevier issued an \href{https://www.elsevier.com/about/policies-and-standards/generative-ai-policies-for-journals}{AI‑usage policy} as early as 2024, requiring authors submitting to its journals to include an AI Usage Statement. The dataset includes all articles published in these journals in 2024. A total of 8,633 Elsevier articles  were collected. After manual inspection, 136 journal-source records remained; one explicit no-AI declaration was excluded from positive AI-use analyses, leaving N = 135 positive AI-use declarations. Table \ref{journal} presents the specific scope of journals included in this study.

After determining the selection scope of journals, statements of "Declaration of Generative AI and AI-assisted technologies in the writing process" were extracted from the articles.  These statements describe which AI tools the authors used and specify their intended purposes. An example is as follows:

\textit{“Statement: During the preparation of this work, the author(s) used ChatGPT to generate essays. After using this tool/service, the author(s) reviewed and edited the content as needed and take(s) full responsibility for the content of the publication.”}

In addition, metadata of the articles (title, author names, journal title) and information about the authors' team (authors' background and language information) were collected.

\begin{table}
\centering    

\small
\caption{Scope of journal sources}
\begin{tabular}{p{5cm}p{5cm}p{1cm}p{1cm}} 
\toprule
Major Categories & Journal Title & CiteScore & AI Decl.(\textperthousand)\\
\midrule
Agricultural and Biological Sciences & Environmental Technology \& Innovation & 14.0 & 4.30\\
Art and Humanities & Acta Psychologica & 3.0 & 28.10\\
Biochemistry & eBioMedicine & 17.7 & 8.71\\
Business, Management and Accounting & Digital Business & 7.4 & 80.00\\
Chemical Engineering & Ultrasonics Sonochemistry & 15.8 & 7.37\\
Chemistry & Redox Biology & 19.9 & 11.76\\
Computer Science & Int. J. Information Management Data Insights & 19.2 & 43.50\\
Decision Sciences & Transport. Research Interdisciplinary Persp. & 12.9 & 27.21\\
Dentistry & The Saudi Dental Journal & 3.6 & 3.89\\
Earth and Planetary Sciences & Progress in Disaster Science & 14.6 & 60.97\\
Economics & Journal of Innovation \& Knowledge & 16.1 & 23.26\\
Engineering & Advances in Applied Energy & 23.9 & 73.17\\
Environmental Science & Environment International & 21.9 & 3.68\\
Health Professions & Complementary Therapies in Medicine & 8.6 & 31.25\\
Immunology and Microbiology & J. Microbiology, Immunology and Infection & 15.9 & 23.80\\
Materials Science & Materials \& Design & 14.3 & 15.94\\
Mathematics & Partial Diff. Equations in Applied Math. & 6.2 & 2.40\\
Medicine & Int. J. Infectious Diseases & 18.9 & 1.48\\
Multidisciplinary & Scientific African & 5.6 & 23.06\\
Neuroscience & Brain Stimulation & 7.6 & 26.31\\
Nursing & Journal of Functional Foods & 9.6 & 7.33\\
Pharmacology & Molecular Therapy Nucleic Acids & 15.4 & 3.77\\
Physics and Astronomy & Ultrasonics Sonochemistry & 15.8 & 7.35\\
Psychology & Comprehensive Psychiatry & 12.5 & 9.90\\
Social Sciences & Computers and Education: AI & 16.8 & 201.43\\
Veterinary & Current Research in Parasitology & 3.6 & 15.38\\
\bottomrule
\end{tabular}
\label{journal}
\end{table}

\subsection{Design of the study}
\begin{figure}[h]
    \centering
    \includegraphics[width=0.9\linewidth]{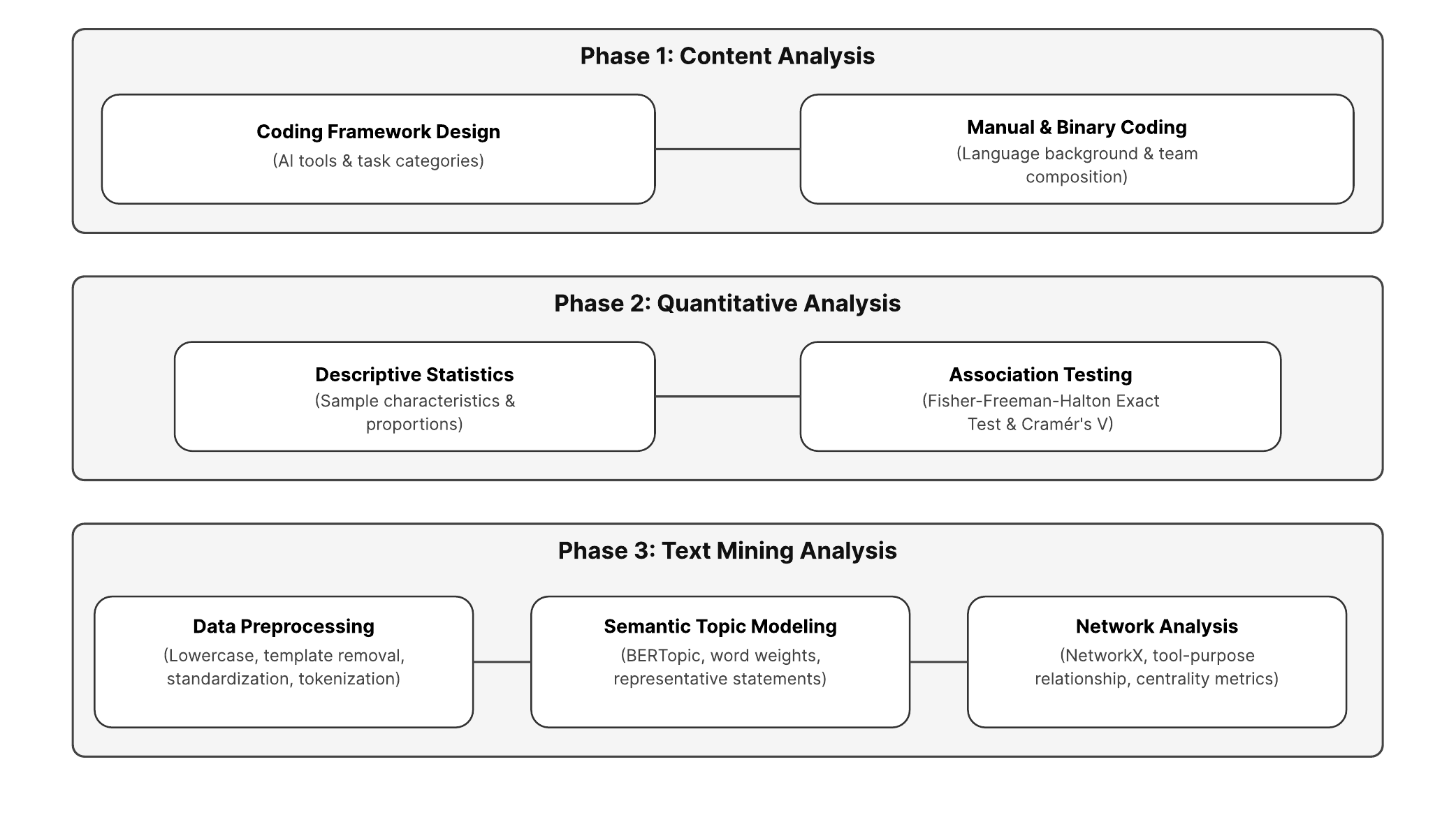}
    \caption{Data Analysis Flowchart}
    \label{fig:Data Analysis Flowchart}
\end{figure}

To conduct this study, an embedded mixed methods design was employed. Specifically, this study used content analysis, quantitative analysis, and text mining methods to systematically examine the use of GAI tools in academic writing, their application tasks and the impact of team characteristics. Figure \ref{fig:Data Analysis Flowchart} provides a detailed illustration of the research design of this study.

\subsubsection{Content Analysis}

To understand the primary AI tools at the declaration level, additional tool mentions, and the tasks for which the tools were declared to be used, this study conducts a systematic content analysis of the collected AI usage declarations.  First, the researchers designed a coding framework to categorize and code the main information in the declarations, including types of AI tools (e.g., ChatGPT, Grammarly) and specific tasks (e.g., grammar checking, sentence polishing, content generation).  Furthermore, in the coding process, binary coding was applied to indicate the first author's language background and team composition, where 0 denotes negative cases and 1 denotes positive cases. 

\subsubsection{Quantitative Analysis}

The quantitative analysis consists of two parts: descriptive statistical analysis and variable association analysis. To address Research Question 3, this paper proposes the following hypotheses: 

\textbf{Hypothesis 1: There is a significant association between authors' native language status and their purposes for using AI tools.}

Non-native English speaking international graduate students face challenges in academic writing due to English being a second language \citep{Singh2015InternationalGS,Flowerdew2019TheLD}.  Artificial intelligence can help non-native English-speaking scientists overcome challenges in scientific writing \citep{giglio2023use}. Writing assistance tools for ESL learners have been developed to address various challenges faced by non-native English writers. These tools focus on error correction \citep{leacock2009user}, planning and writing support \citep{lim2012esl}, and first-language-oriented assistance \citep{chen2012flow}.

\textbf{Hypothesis 2: There is a significant association between team composition and their purposes for using AI tools. }

 Research on communication in international collaborative research teams remains limited, with gaps in understanding team diversity, language use, and communication processes in the social sciences and humanities \citep{wohlert2020communication}. 
 However, information and communication technologies (ICT) can mitigate these negative effects on intercultural communication while supporting the positive impact on decision-making \citep{Shachaf2008CulturalDA}. International teams tend to use richer communication media more intensively than domestic teams when dealing with complex tasks \citep{Bjorvatn2019ComplexityAA}. In collaborative writing, mixed nationality pairs focus more on content discussion, while same-nationality pairs engage more in language-related aspects \citep{rahayu2020interaction}.

The dataset was categorized using two indicators: language background and team composition characteristics. The final  AI-use sample consisted of N = 135 positive AI-use declarations. Nine descriptive purpose categories were collapsed into six inferential categories: Readability, Grammar, Proofreading, Statistical Analysis, Translation, and Other. The analysis is based on 2×6 contingency tables that include two binary classification grouping variables (Native Speaker/Non-Native Speaker; International Group/Non-International Group) and six usage purpose categories. 

This paper employed the Fisher-Freeman-Halton exact test to analyze the association between these variables, because more than 20\% of cells have expected frequencies less than 5, violating chi-square test assumptions. This test serves as an extension of Fisher's exact test on RxC contingency tables, particularly suitable for analyzing categorical data of this type with small samples. To avoid relying only on statistical significance, Cramér’s V was reported as an effect size measure for each contingency table.

\subsubsection{Text Mining}

To enhance analytical efficiency and support content analysis, this study further employs text mining techniques to automate the analysis of declaration texts. This section consists of two main parts: basic textual analysis and semantic pattern analysis. The text analysis process comprises three main steps:

(1) Data preprocessing: Initially, all text was converted to lowercase. Second, template sentences from the original declaration statements were removed. Third, term standardization was performed to address inconsistencies in authors' expression of technical terms (e.g., converting 'gpt-4' to 'chatgpt4'). Finally, tokenization and stop word removal were conducted.


(2) Semantic analysis: The BERTopic was employed to identify latent topics in the disclosure corpus. For the each topic, I inspected the the highest-weighted terms and representative statements. BERTopic was selected because it combines transformer-based document embeddings with class-based TF-IDF topic representation, making it suitable for short and formulaic disclosure texts

(3) Relationship extraction and network analysis: In the relationship extraction phase, Python's NetworkX package was used to extract tool-purpose relationships and tool-tool collaborative relationships, with frequency weights assigned to each relationship level. During the network construction phase, graph structures were built with nodes and edges added. The network analysis phase involved calculating network statistics and centrality metrics. Finally, data visualization was performed using undirected bipartite weighted network graphs.

\section{Results}

\subsection{Content Analysis Results}

\begin{figure}[ht]
    \centering
    \includegraphics[width=\linewidth]{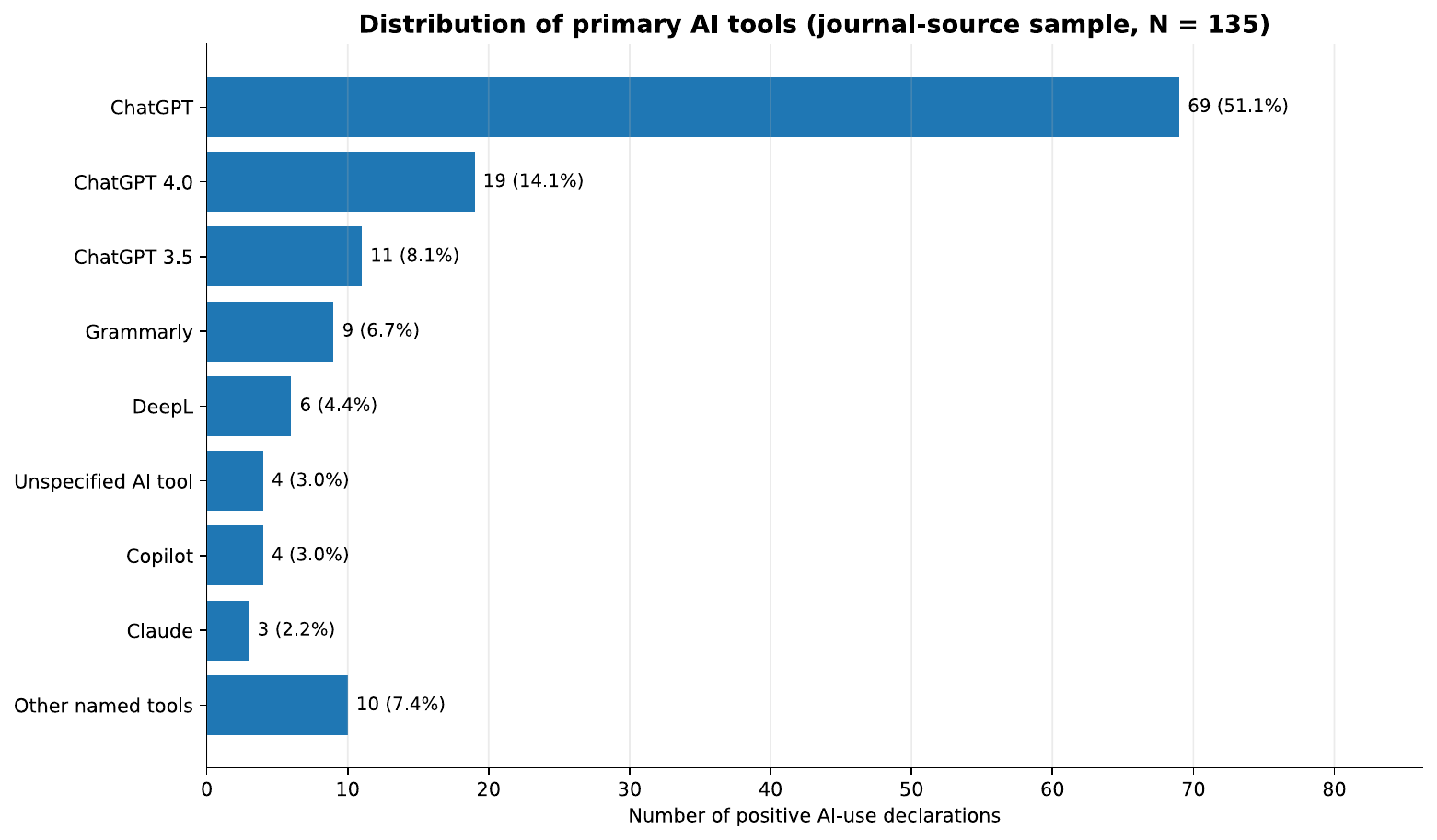}
    \caption{AI tools usage distribution}
    \label{fig1}
\end{figure}

To provide an overview of the AI tools reported in disclosure statements, Figure \ref{fig1} presents the distribution of declared AI tools. At the declaration level, ChatGPT-family tools were the dominant primary tools, accounting for 99 of 135 positive AI-use declarations (73.3\%). The majority of researchers only emphasize the application of ChatGPT, without declaring the specific version. Furthermore, Grammarly and DeepL occupy the fourth and fifth positions respectively, due to their specialized capabilities in grammar-checking and translation. 

\begin{figure}[H]
    \centering
    \includegraphics[width=\linewidth]{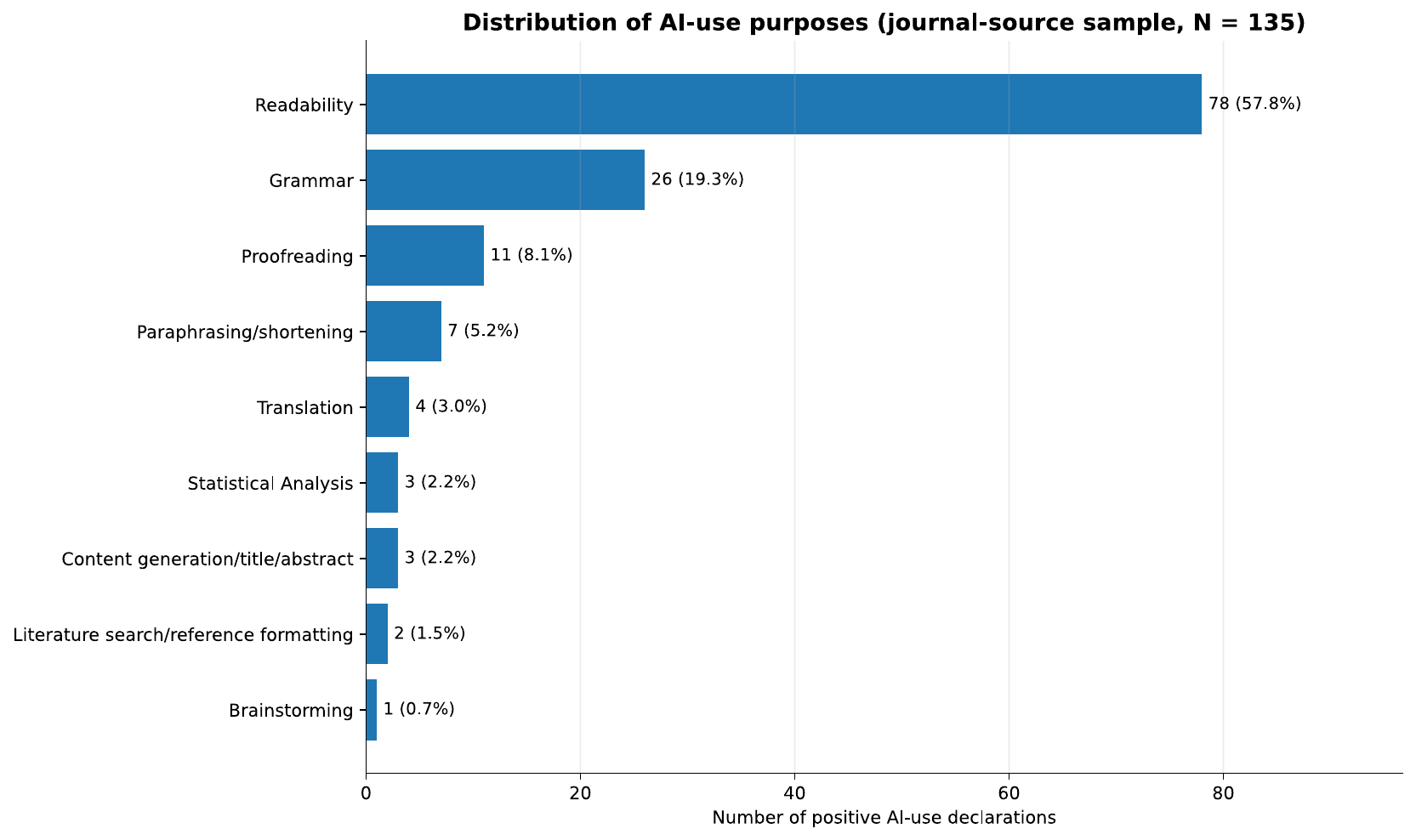}
    \caption{Distribution of AI tools Usage Purpose}
    \label{ai_tools_purpose_pie}
\end{figure}
To better understand the functional roles of AI tools in academic writing, Figure \ref{ai_tools_purpose_pie} summarizes the distribution of declared usage purposes across the dataset. 
As shown in Figure \ref{ai_tools_purpose_pie}, the purposes of AI tools usage are classified in nine categories, ranging from readability, grammar, proofreading, statistical analysis, translation, paraphrasing, generating titles or abstracts, searching literature, and brainstorming. Representative examples of declaration statements for each category are provided in Table \ref{tab:purpose_examples}. The main reason that authors use AI tools is to improve the readability of manuscripts, accounting for 57.8\%. Grammar-checking is the second most common purpose, with 19.3\% of declarations. It is noteworthy that some scholars utilize AI tools for statistical analysis and content generation. While these various applications primarily serve auxiliary functions during the academic writing process, researchers also employ tools like ChatGPT for brainstorming and literature searches in the preliminary stages of academic writing. In addition, the utilization of AI tools in academic writing is not limited to a single application. Twenty-one percent of scholars employed a combination of two different AI tools, while a minority of researchers utilized three tools to facilitate their academic writing process. 

\begin{table}[ht]
\centering
\caption{Examples of AI Tool Usage Purposes in Academic Writing}
\begin{tabular}{p{1.5cm}p{12cm}}
\toprule
Purpose & Example Statement \\
\midrule
Readability & The authors utilized GPT-4 from OpenAI to enhance language and readability. \\
\addlinespace
Grammar & During the preparation of this work the authors used Grammarly, Inc. in order to review spelling and grammar. \\
\addlinespace
Proofreading & We employed ChatGPT 4.0 for proofreading the text of this contribution. \\
\addlinespace
Statistical Analysis & For the preparation of this paper, a proprietary large language model, i.e., GPT-4 Turbo, was utilized specifically in the process of labeling the detected communities within our dataset. \\
\addlinespace
Translation & During the preparation of this work, the authors used DeepL to translate parts of the manuscript from German to English. \\
\addlinespace
Paraphrase & In preparing this work, the author(s) employed the assistance of ChatGPT 3.5 to paraphrase the manuscript. \\
\addlinespace
Content Generation & The authors used ChatGPT solely for English proofreading, and the graphical abstract was partially designed by DALL-E 3. \\
\addlinespace
Brainstorming & ChatGPT 4 and GrammarlyGO were used to brainstorm and refine sentence structures in select parts of the manuscript. \\
\addlinespace
Literature Search & The author(s) used Microsoft's Bing AI to search for literature and format references. \\
\bottomrule
\multicolumn{2}{l}{\small{Note: These statements are representative examples extracted from published papers.}}
\end{tabular}
\label{tab:purpose_examples}
\end{table}

\subsection{Quantitative Analysis Results}
Before examining associations between author characteristics and AI-use purposes, Table \ref{tab:distribution} presents the overall distribution of language background and team composition variables. Non-native speakers constitute the vast majority of the sample, accounting for 83.0\%. Regarding team composition distribution, more than two-thirds of the declarations were from non-international teams. To further illustrate the sample composition and prepare for association analysis, Table \ref{tab:cross_analysis} presents a cross-analysis of language background and team characteristics. The vast majority of AI usage declarations originate from non-native speakers working in non-international teams (59.3\%). This suggests that researchers lacking both native English proficiency and the immediate linguistic support of international collaborators represent the primary demographic relying on AI assistance. Conversely, native speakers constitute a small minority, with native speakers in non-international teams forming the smallest subgroup (8.1\%).

Table \ref{tab:tool_usage} displays AI tool usage patterns among authors with different backgrounds. Non-native speakers exhibit a broader diversification in tool selection, most notably relying on specialized translation software like DeepL, which is entirely absent among native speakers. Furthermore, native speakers show a proportionally notable reliance on Grammarly relative to their small sample size, indicating that their use of AI is primarily geared toward stylistic refinement and proofreading rather than foundational translation or content generation.

\begin{table}[htbp]
\centering
\caption{Distribution of Author Characteristics}
\begin{tabular}{lcc}
\toprule
Characteristics & Count & Percentage (\%) \\
\midrule
Native Speaker & 23 & 17.0 \\
Non-native Speaker & 112 & 83.0 \\
\midrule
International Group & 44 & 32.6 \\
Non-international Group & 91 & 67.4 \\
\bottomrule
\end{tabular}
\label{tab:distribution}
\end{table}

\begin{table}[htbp]
\centering
\caption{Cross Analysis of Author Background}
\begin{tabular}{lccc}
\toprule
Native Speaker & International Group & Count & Percentage (\%) \\
\midrule
Yes & Yes & 12 & 8.9 \\
Yes & No & 11 & 8.1 \\
No & Yes & 32 & 23.7 \\
No & No & 80 & 59.3 \\
\bottomrule
\end{tabular}
\label{tab:cross_analysis}
\end{table}

\begin{table}[htbp]
\centering
\caption{AI Tools Usage by Author Background}
\begin{tabular}{lcccc}
\toprule
\multirow{2}{*}{AI Tools} & \multicolumn{2}{c}{Native Speaker} & \multicolumn{2}{c}{International Group} \\
\cmidrule(lr){2-3} \cmidrule(lr){4-5}
& Yes & No & Yes & No \\
\midrule
ChatGPT & 10 & 59 & 22 & 47 \\
ChatGPT 4.0 & 5 & 14 & 7 & 12 \\
ChatGPT 3.5 & 2 & 9 & 4 & 7 \\
Grammarly & 3 & 6 & 4 & 5 \\
DeepL & 0 & 6 & 1 & 5 \\
Unspecified AI tool & 0 & 4 & 1 & 3 \\
Copilot & 1 & 3 & 3 & 1 \\
Claude & 0 & 1 & 1 & 2 \\
Other named tools & 2 & 8 & 1 & 9 \\
\bottomrule
\end{tabular}
\label{tab:tool_usage}
\end{table}

\textbf{Analysis of Authors' Language Background (H1)} 

Table \ref{tab:purpose_association} shows the association between author background variables and disclosed AI-use purposes. For native-speaker status, the Fisher-Freeman-Halton exact test was not statistically significant, p = 0.2359, and the effect size was weak, Cramér’s V = 0.219. Therefore, H1 was not supported. Despite the lack of overall statistical significance, descriptive data reveal that both native and non-native speakers primarily use AI tools for readability (69.6\% and 55.4\%, respectively). While non-native speakers showed a higher proportional use for grammar correction (22.3\% vs 4.3\%) and were the only group to report using AI for translation, these observational differences did not reach statistical significance.

\textbf{Team Composition Analysis (H2)} 

The results of the Fisher-Freeman-Halton exact test show a highly significant association between team composition and the purpose of using AI tools ($p = 0.0008 < 0.01$),with a moderate effect size, Cramér’s V = 0.382. Therefore, H2 was supported. Key findings indicate that international teams place significantly more emphasis on grammar compared to non-international teams (31.8\% vs 13.2\%). Additionally, no proofreading, statistical analysis, or translation purposes were observed among international teams. Conversely, the proportion of international teams utilizing AI for "other" purposes is notably higher than that of non-international teams (18.2\% vs 5.5\%).

\begin{table}[htbp]
\small 
\centering
\caption{Association between Author Background and AI-Use Purpose Categories with Effect Sizes}
\begin{tabular}{lcccccc}
\toprule
\multicolumn{7}{l}{\textbf{Panel A: Native Speaker Status}} \\
\midrule
& Read. & Gram. & Proof. & Anal. & Trans. & Other \\
\midrule
Yes & 16 & 1 & 3 & 0 & 0 & 3 \\
(n=23) & (69.6\%) & (4.3\%) & (13.0\%) & (0.0\%) & (0.0\%) & (13.0\%) \\
\addlinespace
No & 62 & 25 & 8 & 3 & 4 & 10 \\
(n=112) & (55.4\%) & (22.3\%) & (7.1\%) & (2.7\%) & (3.6\%) & (8.9\%) \\
\midrule
\multicolumn{7}{l}{Fisher-Freeman-Halton test: p = 0.2359; Cramér’s V = 0.219.} \\
\midrule
\multicolumn{7}{l}{\textbf{Panel B: International Group Status}} \\
\midrule
& Read. & Gram. & Proof. & Anal. & Trans. & Other \\
\midrule
Yes & 22 & 14 & 0 & 0 & 0 & 8 \\
(n=44) & (50.0\%) & (31.8\%) & (0.0\%) & (0.0\%) & (0.0\%) & (18.2\%) \\
\addlinespace
No & 56 & 12 & 11 & 3 & 4 & 5 \\
(n=91) & (61.5\%) & (13.2\%) & (12.1\%) & (3.3\%) & (4.4\%) & (5.5\%) \\
\midrule
\multicolumn{7}{l}{Fisher-Freeman-Halton test: p = 0.0008**; Cramér’s V = 0.382} \\
\bottomrule
\multicolumn{7}{p{0.95\linewidth}}{\small{Note: **p \textless 0.01. Read. = Readability, Gram. = Grammar, }} \\
\multicolumn{7}{p{0.95\linewidth}}{\small{Proof. = Proofreading, Anal. = Statistical Analysis, Trans. = Translation.}} \\
\multicolumn{7}{p{0.95\linewidth}}{\small{Percentages in parentheses represent within-group proportions.}}
\end{tabular}
\label{tab:purpose_association}
\end{table}

\subsection{Text Mining Results}
To uncover the latent thematic structures within the authors' self-disclosed AI purposes, BERTopic modeling was applied to the corpus. As shown in Table \ref{tab:semantic_clusters}, BERTopic identified nine non-outlier topics. The topics were overwhelmingly language-related, including general readability and language assistance, grammar and spelling checking, and fluency or coherence enhancement. These results indicate that disclosed AI use was primarily oriented toward linguistic support rather than substantive research generation.

\begin{table}[ht]
\centering
\begin{tabular}{lrr p{4cm} p{5cm}}
\hline
\textbf{Topic} & \textbf{n} & \textbf{\%} & \textbf{Label} & \textbf{Representative terms} \\
\hline
T0  & 33 & 24.4\% & General readability and language assistance & readability, language, improve, assist \\
T1  & 22 & 16.3\% & Grammar and spelling checking & grammar, check, spelling, wording, structure \\
T2  & 20 & 14.8\% & Language readability improvement & language readability, improve language, English, sentences \\
T3  & 17 & 12.6\% & Readability and language polishing & readability, language, streamline, clarity \\
T4  & 14 & 10.4\% & English proofreading and copyediting & English, editing, streamline, correct grammar \\
T5  &  9 &  6.7\% & Clarity-oriented sentence revision & clarity, revise, sentences, structure \\
T6  &  7 &  5.2\% & Section-level streamlining & sections, text, introduction, certain sections \\
T7  &  6 &  4.4\% & English-language polishing and flow & English, language, text, polish, linguistic \\
T8  &  5 &  3.7\% & Fluency and coherence enhancement & fluency, coherence, quality, language \\
T-1 &  2 &  1.5\% & Outliers & --- \\
\hline
\end{tabular}
\caption{Semantic clusters of purpose-bearing clauses.}
\label{tab:semantic_clusters}
\end{table}

\begin{figure}[h]
    \centering
    \includegraphics[width=1\linewidth]{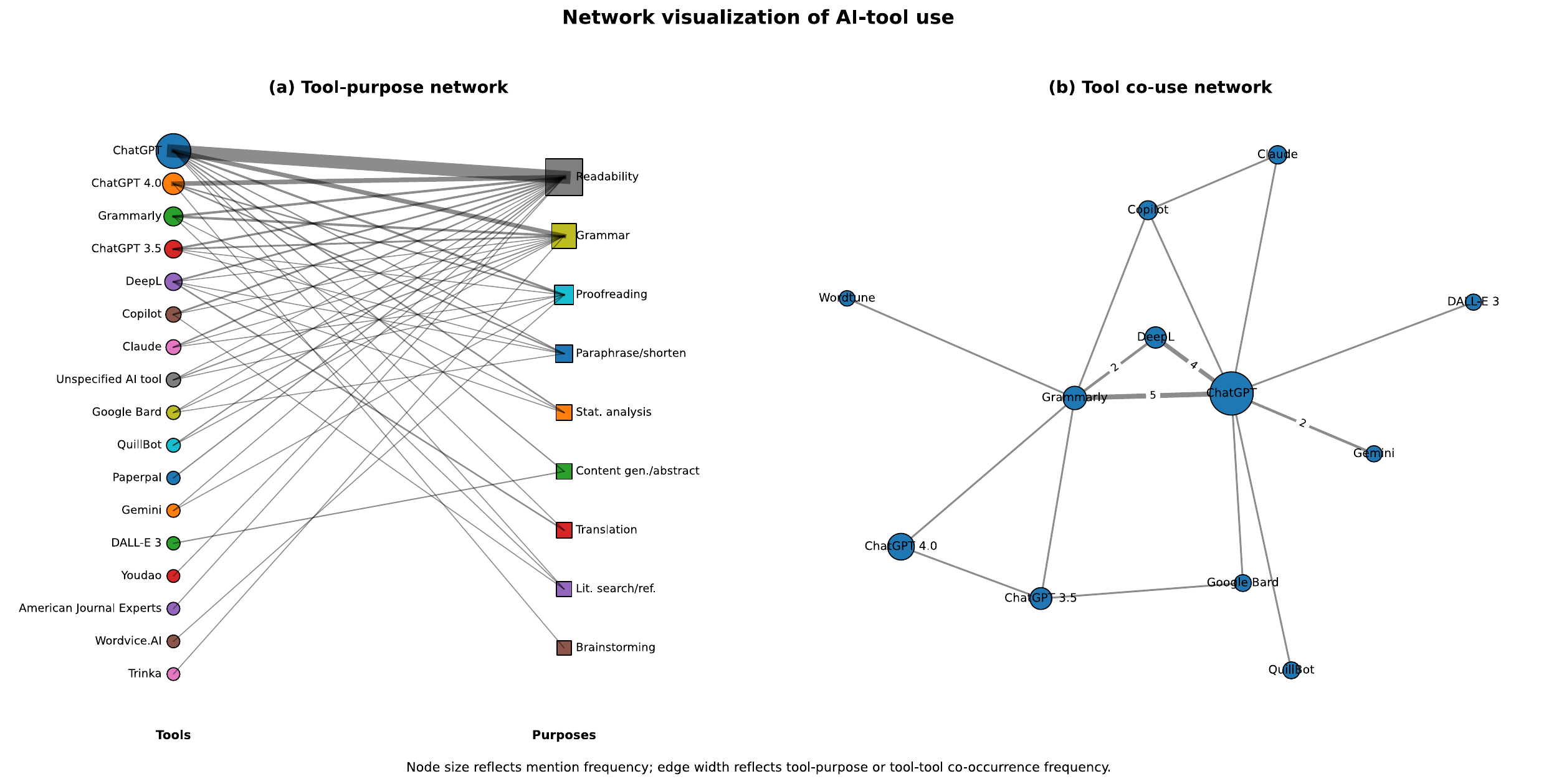}
    \caption{Network Visualization}
    \label{fig:network_visualization}
\end{figure}

As mentioned above, some authors utilize multiple AI tools to assist in academic writing. Through semantic analysis, ChatGPT demonstrates the highest centrality in the semantic network (see Figure \ref{fig:network_visualization}b). ChatGPT shows strong connections with its other versions (4.0, 3.5). Grammarly is frequently used in conjunction with the ChatGPT family of tools. DeepL and Claude tend to be used as supplementary tools.
In the tool-purpose network graph (Figure \ref{fig:network_visualization}a), the ChatGPT family of tools is primarily employed for improving readability, while Grammarly focuses on grammar checking. DeepL and Claude serve more specialized purposes.

\section{Discussion}
This study aimed to explore the trend and purpose of AI tools used by researchers in academic writing. In addition, this paper examined the association of authors' background and AI tools usage. Three notable findings emerged in this study: (a) Dominance and Diversity in Adoption of AI Tools, (b) Author Background Influence on AI Tool Utilization and (c) Primary Applications and Emerging Use Cases. 

Firstly, ChatGPT family tools are predominant AI tools in academic writing.  This finding is corroborated by other studies. \citet{selim2024transformative}'s research on non-native English-speaking university students revealed that ChatGPT and Grammarly are the two most frequently used AI tools. 
The emergence of ChatGPT, as a pioneering generative artificial intelligence system, transformed conventional understanding of AI agents and generated extensive societal discussions. This first-mover advantage has contributed to ChatGPT becoming the most extensively utilized large language model. However, researchers, especially non-native speakers, also utilized other AI tools, such as DeepL and Grammarly, which had a huge number of users prior to the prevalence of ChatGPT, due to user loyalty and the support of LLMs.

According to the quantitative findings, both native and non‑native speakers most frequently used AI to enhance text readability. This suggests that using AI to support writing has become a shared need, as even native English speakers rely on AI to improve the fluency and logical coherence of their texts. Although $H_1$ was not supported, the descriptive data clearly reveal the unique challenges faced by non‑native speakers: only non‑native users reported using AI for translation, and they also showed a higher reliance on AI for grammar‑related assistance.

For non-native English-speaking researchers, AI tools can help translate literature to support their academic research \citep{zenni2023artificial,hwang2023chatgpt,song2023enhancing}. 
The highly significant association found in $H_2$ highlights the distinct role AI plays in collaborative settings. International groups prefer to utilize Grammarly more than the non-international group, which indicates specific challenges in cross-language communication. The unification of writing style through these tools leads to reduced communication costs arising from language differences.

To provide differentiated support for researchers, academic journals are required to develop targeted guidelines for the use of AI tools. However, numerous academic journals lack established AI usage policies, not to mention differentiated support systems.  According to a survey of the top 100 academic publishers, only 24\%  provided guidance on the use of Generative AI \citep{ganjavi2024publishers}.

Thirdly, improving manuscript readability is the primary purpose, while grammar checking is the secondary purpose. Additionally, statistical analysis and literature search/reference 
formatting were less frequent but notable use cases. This raises new academic ethical questions about whether non-language enhancement purposes, such as data analysis and abstract generation, should be permitted in AI tool usage. The scope of AI usage in academic contexts remains an unresolved issue.

Some academic journals encourage authors to use AI for editing manuscript, provided that the technology’s contribution is appropriately recognized \citep{crawford2023artificial}.  However, the specific effectiveness and extent of policy implementation remain uncertain, considering the accuracy of AI-generated text \citep{Hu06122024}. Academics across disciplines have been quick to adopt new technologies and tools to assist with research and practice \citep{GREWAL2021229}.  The ongoing discussions include considerations about listing ChatGPT as a co-author, requirements for disclosing AI usage in the acknowledgments section, and the potential necessity of submitting a separate declaration document to journals specifying AI-generated content \citep{doi:10.1177/14413582231167882}. 

Finally, a notable finding reveals that researchers integrate AI tools throughout the entire research cycle. At the pre-research stage, generative AI facilitates brainstorming and exploration of diverse research angles. During manuscript development, AI supports data analysis and abstract composition. \citet{huang2024evaluating} argue that researchers tend to use generative AI tools as auxiliary tools to analyze data and inspire more innovative ideas. In the final stage, it aids in refining language and improving readability. ChatGPT can take on the role of a virtual advisor that supports study and research, not only resolving questions but also providing reviews and feedback \citep{choi2021chatgpt}. AI shifts authorship from a purely writing-centered role toward a supervisory and accountability-oriented role.

\subsection{Theoretical Implications}

The current practice of declaring AI usage in academic publishing is emerging as a new transparency norm, conceptually akin to conflict of interest (COI) statements. However, as my findings suggest, these disclosures primarily reflect what authors are willing to acknowledge rather than the full extent of their actual AI integration. According to \cite{HeBu2026AIpolicy}, existing publisher guidelines have not yet effectively regulated or transparently standardized AI use in scientific writing. As publishers increasingly mandate these disclosures, authors are continually adapting to this evolving academic norm. Moving forward, this norm must extend beyond simple declarations of tool names or surface-level purposes. A robust framework will require a granular, transparent delineation of which research components are human-generated versus AI-generated. As discussed in recent literature on governing AI in publishing ecosystems \citep{Frieder}, radical future policy reforms might even necessitate AI systems themselves cryptographically verifying these statements. While such shifts could redistribute academic power and exacerbate inequalities in AI resource access, the rapid evolution of generative AI necessitates dynamic and adaptive regulatory policies that can keep pace with technological advancement \citep{KaraarslanAydin2026AIpublishing}.

Fundamentally, the proliferation of generative AI has blurred the traditional boundaries of authorship and academic misconduct, prompting a necessary reconceptualization of "AI-assisted scholarly labor." The perception of this labor varies significantly across academic stakeholders; for instance, students may view AI as a ubiquitous utility akin to a calculator, whereas educators often express concerns regarding cognitive degradation \citep{10.1108/IJILT-11-2023-0213}. To systematically conceptualize this labor, we can draw upon the taxonomy proposed by the Research Center for Social Computing and Interactive Robotics at Harbin Institute of Technology, which categorizes "AI for Research" into five distinct dimensions: (1) Scientific Comprehension, (2) Academic Surveys, (3) Scientific Discovery, (4) Academic Writing, and (5) Academic Reviewing \citep{chen2025ai4researchsurveyartificialintelligence}. By applying this taxonomy to my findings, it becomes evident that current AI disclosures are heavily concentrated in the fourth dimension: Academic Writing. This practical insight suggests that "AI-assisted scholarly labor" is not a single, uniform activity.

\subsection{Practical Implications
}
Although scholars have called for academic organizations, such as the APA, to generate AI usage guidelines in research, this guide does not focus on different hierarchical requirements \citep{tate2023educational}. As a next step, AI policy is supposed to distinguish between basic language polishing and deep content generation. In order to improve the transparency of AI usage, publishers need to design templates for AI-use disclosures. Many statements merely declare which AI tools were utilized, without specifying the tasks and scopes of application. The review system should not take a one-size-fits-all approach, but rather adopt different evaluation standards for various types of AI applications (such as translation versus content generation).

Researchers exhibit significant variations in their use of AI tools, reflecting the necessity of AI literacy education. Researchers need to master multiple AI tools, which requires comprehensive training support. Therefore, AI literacy should be incorporated into graduate courses. Although AI streamlines the research process, it sets higher standards for critical thinking (the ability to evaluate the reliability of AI outputs) and ethical awareness (understanding the scope and responsibilities of AI). ChatGPT usually provides fabricated references \citep{Mjovsk2023ArtificialIC}. AI literacy should not be confined to the writing process. According to our findings, AI tools span the entire research cycle. The education of AI literacy needs to break from traditional frameworks and develop a comprehensive framework that addresses AI applications across different research stages, from literature review and data analysis to result interpretation and scholarly communication.

Future research on AI tools should focus on explainable AI. LLMs cannot independently produce novel academic papers and research reports. Considering research integrity and rigor, researchers must identify the sources of AI-generated content when citing sentences produced by AI systems. To achieve this objective, AI systems need to enhance their knowledge tracing mechanisms and LLM evaluation frameworks.

Another critical research direction is the integration of diverse functionalities. The fragmented nature of current tools increases operational overhead while reducing efficiency. While major technology and internet corporations are engaged in competition regarding LLM algorithmic performance, they have neglected human-centered development principles and methodologies. The development of AI tools should prioritize interoperability among different systems, facilitating seamless integration of various functions to provide comprehensive solutions. For instance, functionalities such as proofreading, content collaboration, and terminology management should be unified within a single platform. Furthermore, AI tools need to develop discipline-specific functions tailored to various academic contexts and writing scenarios.

\section{Conclusion}

This study reveals several key patterns in the use of AI tools in academic writing. The clear dominance of the ChatGPT family underscores its widespread adoption across diverse scholarly domains. Crucially, while authors' native language status does not significantly affect AI-use purposes, team composition plays a pivotal role. International teams demonstrate a highly significant reliance on AI for grammar and stylistic standardization. The findings highlight the need for more nuanced AI usage guidelines in academic publishing, particularly considering the diverse needs of international research teams. While AI tools primarily serve to enhance readability and grammar, their expanding role in content generation and analysis raises new questions about academic integrity and authorship. 

The present study has certain limitations that should be acknowledged. The primary limitation is the relatively small size of the dataset used in the analysis. Owing to constraints in research capacity, only one journal per discipline was included. Furthermore, due to copyright considerations, the sample was restricted to open‑access journals. Additionally, the reliability of AI usage declarations poses another limitation, as authors' disclosures about their use of AI tools may not always be complete or accurate. It is possible that some authors utilized AI tools in their work without making the appropriate declarations. Future research should use larger datasets, focusing on evolution and current landscape of AI tool usage across different academic disciplines.

\backmatter

\section*{Statements and Declarations}

\subsection*{Conflict of interest/Competing interests}
The author declares that the research was conducted in the absence of any commercial or financial relationships that could be construed as potential conflicts of interest

\subsection*{Data availability}

The datasets generated during and/or analysed during the current study are available from the corresponding author on reasonable request.

\bibliography{references}

\end{document}